# Ferromagnetic order in aged Co-doped $TiO_2$ anatase nanopowders

A.J. Silvestre[1*], L.C.J. Pereira[2], M.R. Nunes[3] and O.C. Monteiro[3]

[1]Instituto Superior de Engenharia de Lisboa and ICEMS, R. Conselheiro Emídio Navarro 1, 1959-007 Lisboa, Portugal.

[2]Departamento de Química, Instituto Tecnológico e Nuclear and CFMCUL, EN10, 2686-953 Sacavém, Portugal

[3]Departamento de Química e Bioquímica, Faculdade de Ciências, Universidade de Lisboa, Campo Grande, 1749-016 Lisboa, Portugal.

## ABSTRACT

Oxide based diluted magnetic semiconductor (DMS) materials have been a subject of increasing interest due to reports of room temperature ferromagnetism in several systems and their potential use in the development of spintronic devices. However, concerns on the stability of the magnetic properties of different DMS systems have been raised. Their magnetic moment is often unstable, vanishing with a characteristic decay time of weeks or months, which precludes the development of real applications.

This paper reports on the ferromagnetic properties of two-year-aged $Ti_{1-x}Co_xO_{2-\delta}$ reduced anatase nanopowders with different Co contents ($0.03 \leq x \leq 0.10$). Aged samples retain rather high values of magnetization, remanence and coercivity which provide strong evidence for a quite preserved long-range ferromagnetic order. In what concern Co segregation, some degree of metastability of the diluted Co doped anatase structure could be inferred in the case of the sample with the higher Co content.

---

* Author to whom correspondence should be addressed. Telephone: +351218317097, Fax: +351218317138, Email: asilvestre@deq.isel.ipl.pt

## 1. INTRODUCTION

Reports of ferromagnetic (FM) order well above room temperature (RT) in the Co-doped $TiO_2$ system[1,2] have generated an increasing interest in diluted magnetic semiconductors (DMS) based on $TiO_2$-doped transition metals because of their potential use in the development of spintronic devices.[3,4] Since then, the synthesis of both anatase and rutile Co:$TiO_2$ ferromagnetic films/nanopowders was achieved using several physical and chemical techniques.[4-15] Magnetic moments of such materials ranging from 0.16 $\mu_B$/Co to values as high as 2 $\mu_B$/Co have been reported.[4,16] Such a wide spread of magnetic moments has raised concerns about the intrinsic nature of the FM properties of the Co:$TiO_2$ films, namely owing to the possibility of cobalt secondary phases,[17-20] heterogeneities or even contamination.[3,21] Although the mechanism by which the ferromagnetism is promoted in Co-doped $TiO_2$ has not yet been definitively clarified and remains the subject of intense debate, we have recently shown that oxygen vacancies play a fundamental role in promoting the long-range FM order in the bulk Co:$TiO_2$ anatase phase, in addition to transition-metal doping.[15] Moreover, our results ruled out the premise of a strict connection between Co clustering and the ferromagnetism reported for this oxide. Altogether, these results seem to be consistent with a FM order explained within the scope of the bound magnetic polaron (BMP) theory.[22-24] Accordingly, when the concentration of shallow defects exceed the percolation threshold, defects such as oxygen vacancies can overlap many dopant ions to yield BMPs, which can result in FM coupling between dopant spins. On the other hand, concerns on the stability of the magnetic properties of different DMS systems have also been raised.[14,25,26] Their magnetic moment is often unstable over time, tending to decay if the same sample is measured on successive occasions, and vanishing with a characteristic decay time of weeks or months. The loss of FM order in these compounds is a problem of major relevance and must be overcome before real industrial applications can be developed.

In this work we investigate the stability of the ferromagnetic order of highly pure Co-doped $TiO_2$ anatase nanopowders, by studying the magnetic properties of two-year-aged $Ti_{1-x}Co_xO_{2-\delta}$ reduced nanopowder samples and comparing them with the magnetic properties of the same samples measured just after preparation (non-aged reduced samples).

## 2. EXPERIMENTAL

Anatase Co:$TiO_2$ nanoparticles with grain sizes in the range 20-30 nm and doping concentrations up to 10 at.% were synthesised near equilibrium conditions by the hydrothermal process previously described by our group.[27] This study is focused on samples with four different Co contents ($x$=0.03, 0.07, 0.08 and 0.10). Cobalt concentrations were determined by coupled plasma-optical emission spectrometry (ICP-OES). $Ti_{1-x}Co_xO_{2-\delta}$ samples were annealed at 500 ºC for 3 h in a reducing atmosphere ($N_2$ + 5% $H_2$) at atmospheric pressure. Details of the microstructure, crystallographic structure, and optical properties of the samples have been published elsewhere.[27] Their magnetic properties were analysed just after preparation in 2008[15] and reanalysed in 2010 after storage at ambient conditions, subject to the natural variations of temperature, pressure and humidity. Magnetization measurements were performed on free powder samples using an Oxford Instruments multicharacterization system *MagLab 2000*. Isothermal magnetization curves were obtained for fields up to 3 T and for temperatures between 4 and 300 K. For all samples the magnetization was measured as a function of temperature in increasing temperature range 5-300 K. The in-phase, $\chi'$, and out-of-phase, $\chi''$, linear components of the AC susceptibility, $\chi_{AC}$, were measured at five different frequencies, from 95 to 9995 Hz in the 2-300 K temperature range, with an AC driving field of 1 Oe.

## 3. RESULTS AND DISCUSSION

DC magnetization *vs.* temperature (*M-T*) curves obtained at a constant applied field of 0.5 T

for the aged Co:$TiO_2$ samples are shown in figure 1. All samples exhibit $M$>0 at RT, the $M$-$T$ curves seem to result from a superposition of a paramagnetic component at low temperature and a ferromagnetic-like one, the rather flat contribution suggesting that the ferromagnetic component has a Curie temperature considerably higher than 300 K, as obtained previously for the non-aged samples.[15] No obvious relation between $M$ and Co content was found, probably due to different oxygen vacancy concentrations induced in the samples, which are difficult to control using post-annealing processes. The isothermal magnetization $M$-$B$ curves obtained at several temperatures confirmed the FM behaviour of the 2 year-aged-reduced $Ti_{1-x}Co_xO_{2-\delta}$ samples, their magnetization reaching near the saturation magnetization value, $M_s$, at $B$=3 T (not shown). $M_s$ values ranging between 0.79 $\mu_B$/Co ($x$=0.07) and 0.05 $\mu_B$/Co ($x$=0.10) were measured at RT for those samples. Figure 2 shows the $M$-$B$ curves obtained at 300 K for $B$ varying in the range [-0.5, 0.5] T for a $Ti_{0.97}Co_{0.03}O_{2-\delta}$ sample, the hysteresis loops (squareness) being clearly resolved for both the non-aged and 2 year-aged samples and determined by the magnetic reversal mechanisms. A slight decrease of the magnetization was observed for all samples except for the one with $x$=0.07 for which a small increase of 4.7 % of the magnetization was observed. Figure 3 summarises these results, showing $M$ at 300 K as a function of the Co content for both sets of samples. It can be seen that the magnetization of the aged samples follows the same trend as that previously measured for the non-aged samples, with the highest value of $M$ obtained for $x$=0.07 composition: 0.68 $\mu_B$/Co (aged sample). Besides the magnetization, special attention was also given to the remanence ($M_r$) and coercivity ($H_c$) since both parameters are a good indication of the extent of ferromagnetic order of the samples. $M_r$ and $H_c$ as functions of Co content for non- and aged samples are shown in figure 4. Both sets of samples evidence identical trends for $M_r$ and $H_c$, with a small decrease for most aged samples relative to the non-aged ones. Nevertheless, it should be noted that $M_r$ values ranging between 0.044 and 0.096 $\mu_B$/Co and $H_c$ values in the range 366.7 Oe - 494.8 Oe were measured for the aged samples. These rather high values seem to be strong evidence that the long-range FM order previously reported for the non-aged samples[15] was quite preserved after two years of sample ageing. Moreover, the $H_c$ values measured for the aged samples are in the range where small magnetic perturbations due to magnetic noise could not easily flip stored data and corrupt a memory device. On the other hand, they are not so large as to require a strong magnetic field, and hence significant power consumption, either to erase old data or write new data.

Another important step in the present study was to investigate the time stability of the dilution of the dopant element in the $TiO_2$ crystal matrix. One may wonder whether Co tends to segregate and form nanoclusters with sample ageing. As reported previously,[15] for the non-aged samples only that with composition $x$=0.08 showed evidence for the presence of Co aggregates, though a dominant long-range FM ordered state was inferred to prevail over the spin-glass-like behaviour of the Co clusters for that sample. For all other non-aged sample compositions ($x$=0.03, 0.07, 0.10) no Co clusters were found. AC susceptibility measurements were used to study Co clustering in the aged samples. Figure 5 shows the temperature dependence of the in-phase $\chi'(T)$ component of $\chi_{AC}$ for the aged samples, taken at 5 different frequencies (95, 495, 995, 4995 and 9995 Hz). The data for samples with $x$=0.03 (Fig. 5a) and $x$=0.07 (Fig. 5b) clearly show a monotonic decrease of the in-phase $\chi'(T)$ susceptibility component with increasing temperature and no anomaly occurs. These facts corroborate the absence of any cluster formation in these aged samples, despite their strong FM behaviour. In contrast, the sample with $x$=0.08 exhibit a strong frequency dependence of $\chi'(T)$ on temperature, the peak position shifting to higher temperatures and its height decreasing as frequency increases (Fig. 5c see arrow). On the other hand, for the out-of-phase $\chi''(T)$ component both the temperature and peak height increase with increasing frequency (not shown). This behaviour is indicative of a freezing/blocking process around a freezing/blocking temperature, $T_B$, associated with the presence of Co clusters, in accordance with that previously reported for the correspondent non-aged sample.[15] Peaks at $T_B$ around 7 K were measured for the $Ti_{0.92}Co_{0.08}O_{2-\delta}$ aged sample. A similar behaviour, though less prominent, was found for the aged sample with $x$=0.10 (Fig. 5d, see dotted circle), for which, when prepared in 2008, no Co clusters were detected (Fig. 5d, inset). This result seems to support the occurrence of some metastability of the diluted Co doped anatase structure, the ageing process tending to induce Co-rich clustering for higher concentrations of the dopant element. The origin of such metastability may be related with the absorption of oxygen during the ageing process, the loss of the excess oxygen vacancies destabilizing the Co doped reduced anatase structure, leading to Co segregation.[28] To gain a deeper insight into the dynamic behaviour of the freezing/blocking process observed for both $Ti_{0.92}Co_{0.08}O_{2-\delta}$ and $Ti_{0.90}Co_{0.10}O_{2-\delta}$ aged samples, the empirical parameter was calculated[29]

$$\Psi = \frac{\Delta T_B}{T_B \Delta \log_{10}(f)}, \tag{1}$$

which represents the relative shift in $T_B$ per frequency decade. $\Delta T_B$ stands for the range of $T_B$ over the $\Delta \log_{10}(f)$ frequency interval. For the dependence observed in $\chi'$, $\Psi$ is found to be ~0.03 for the $Ti_{0.92}Co_{0.08}O_{2-\delta}$ sample and ~0.04 for the $Ti_{0.9}Co_{0.1}O_{2-\delta}$ one. These values are consistent with a cluster-glass behaviour and consequently confirms the presence of Co-rich aggregates in both aged samples.[29]

## 4. CONCLUSIONS

It was shown here that the RT ferromagnetic order of reduced $Ti_{1-x}Co_xO_{2-\delta}$ anatase nanopowders samples is fairly stable in time, in particular for $x \leq 0.07$ compositions. Aged samples retain rather high values of $M$, $M_r$ and $H_c$, which provides strong evidence that the long-range FM order is well preserved after two years of sample ageing. With regard to Co segregation, some degree of metastability of the diluted Co doped anatase structure could be inferred in the case of the sample with higher Co content ($x$=0.10). In fact, over 2 years this sample evolved from a sample free of Co-rich clusters to one consistent with Co-rich cluster-glass behaviour. This particular result shows that high Co contents ($x$~10 at.%) should be avoided whenever stable ferromagnetic samples are required.

**Acknowledgements**

This work was supported by Fundação para a Ciência e Tecnologia (PTCD/CTM/101033/2008). We thank P.I. Teixeira for a critical reading of the manuscript and fruitful discussions.

**Figure captions**

**Figure 1** - *M-T* curves recorded at a constant applied field of 0.5 T for two-year-aged $Ti_{1-x}Co_xO_{2-\delta}$ reduced samples.

**Figure 2** - Isothermal magnetization recorded at 300 K for the non-aged and two-year-aged $Ti_{0.97}Co_{0.03}O_{2-\delta}$ reduced samples. The inset shows a magnification of the hysteresis loops.

**Figure 3** - Magnetization recorded at 300 K at a constant applied field of 0.5 T for the non-aged and two-year-aged reduced samples as a function of Co content.

**Figure 4** - $M_r$ and $H_c$ values recorded at 300 K for the non-aged and two-year-aged reduced samples as a function of Co content.

**Figure 5** - Temperature dependence of the in-phase $\chi'$ component of the AC susceptibility at various frequencies for the two-year-aged $Ti_{1-x}Co_xO_{2-\delta}$ reduced samples. a) $x$=0.03, b) $x$=0.07, c) $x$=0.08 and d) $x$=0.10. The solid arrow in c) passes trough the peak temperatures and is just a guide for the eye. The inset in d) shows $\chi'$ *vs.* $T$ at various frequencies for the non-aged sample with $x$=0.1.

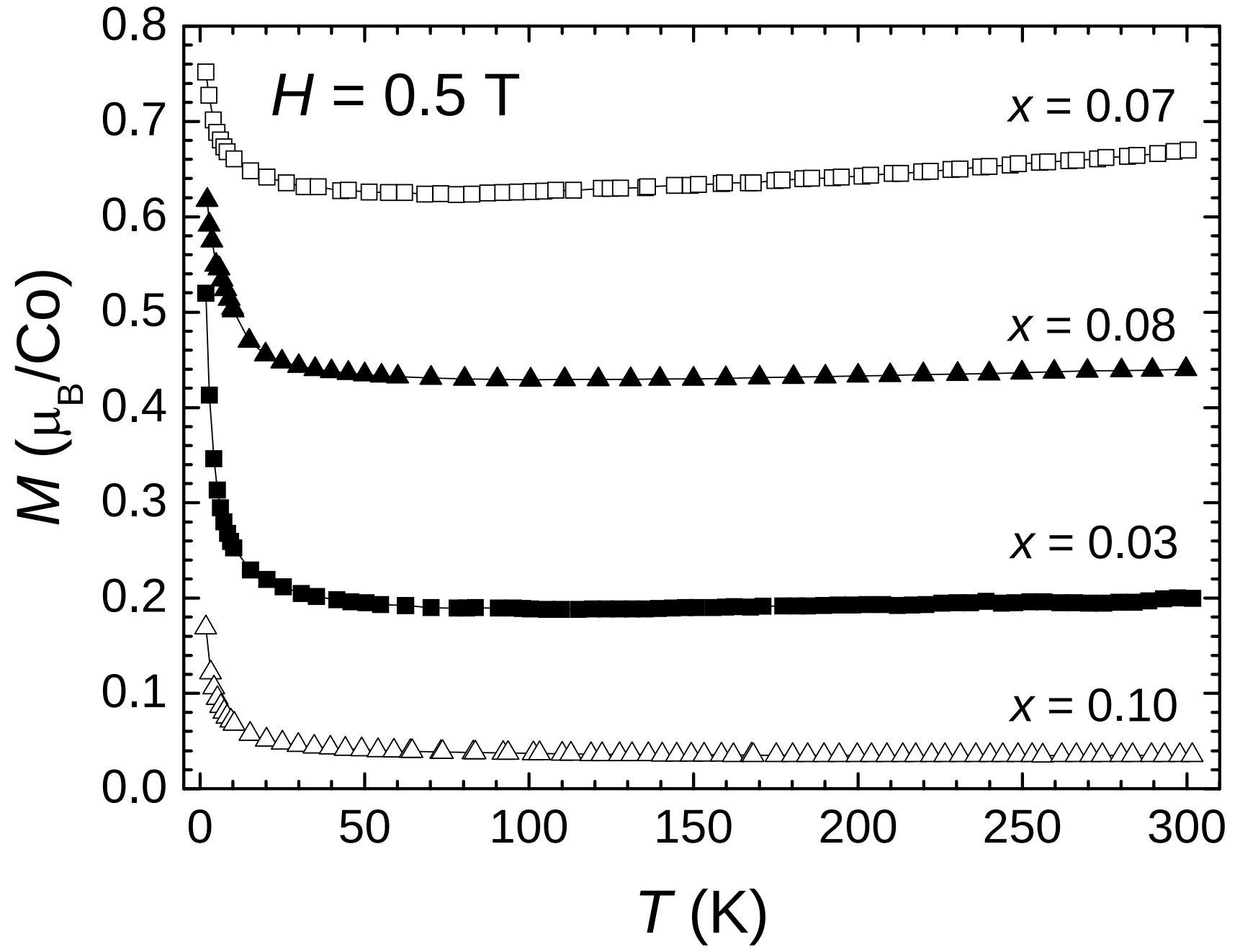

**Figure 1**

**Manuscript title:** Ferromagnetic order in aged Co-doped $TiO_2$ anatase nanopowders

**Authors:** A.J Silvestre *et al.*

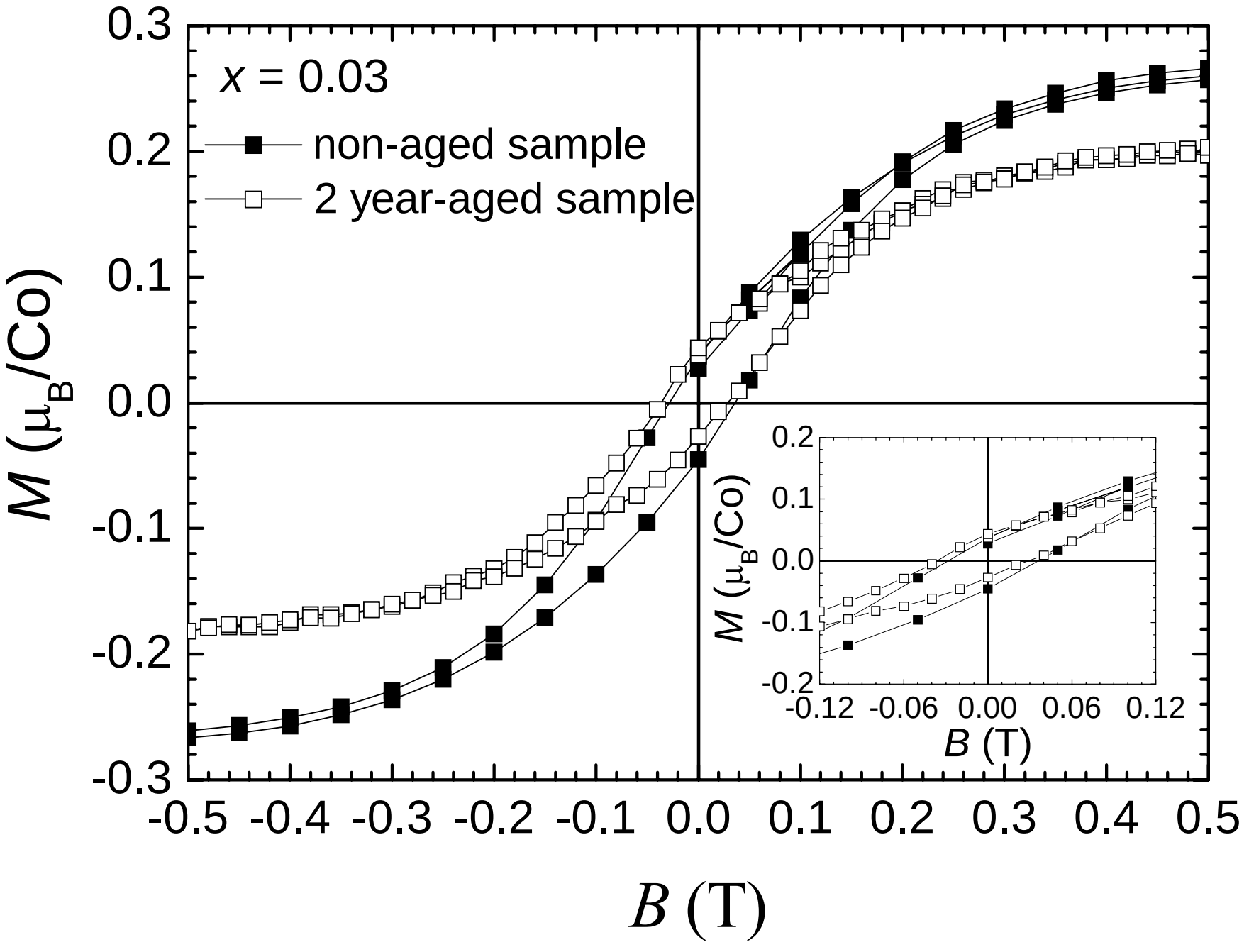

**Figure 2**

**Manuscript title:** Ferromagnetic order in aged Co-doped $TiO_2$ anatase nanopowders

**Authors:** A.J Silvestre *et al.*

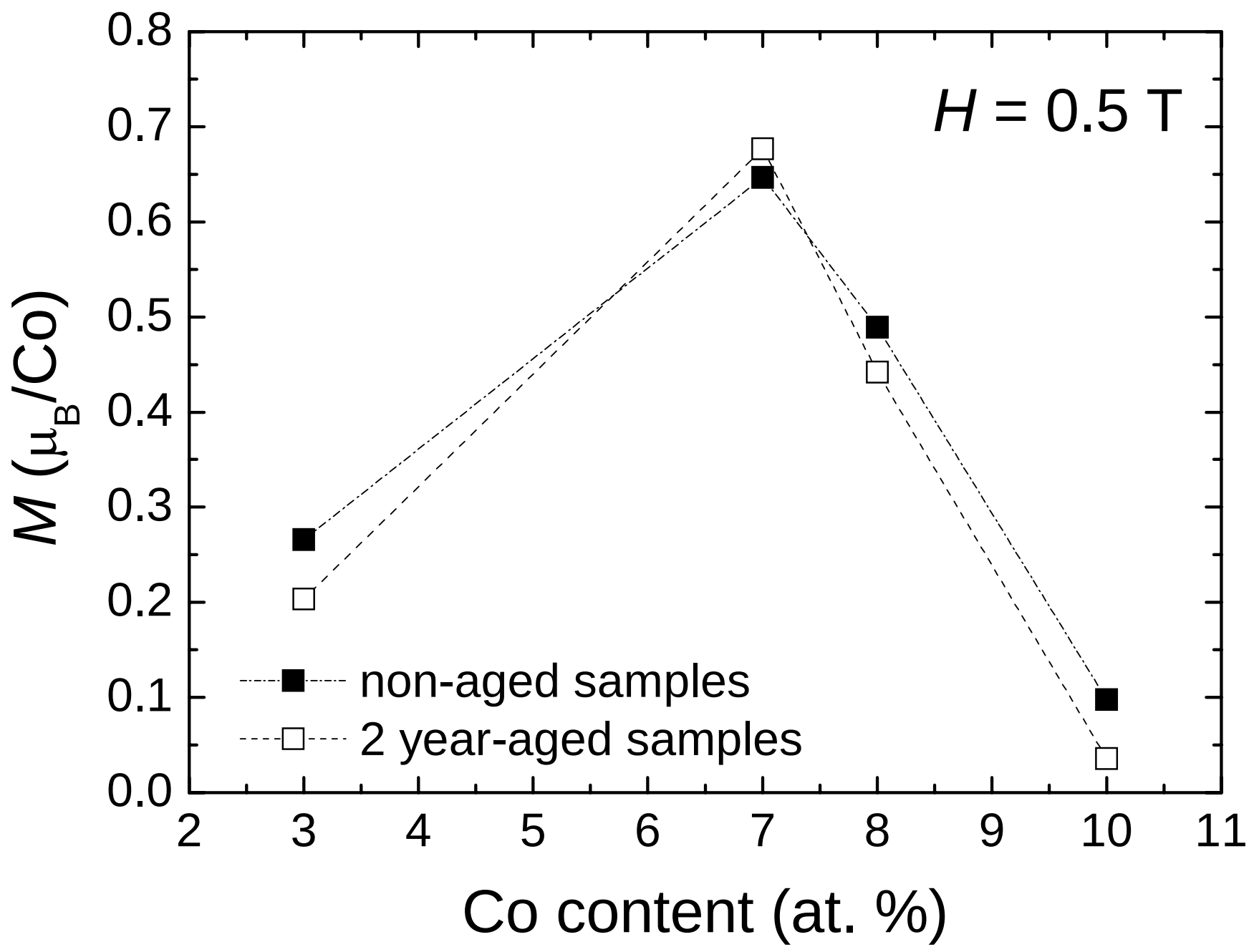

**Figure 3**

**Manuscript title:** Ferromagnetic order in aged Co-doped $TiO_2$ anatase nanopowders

**Authors:** A.J Silvestre *et al.*

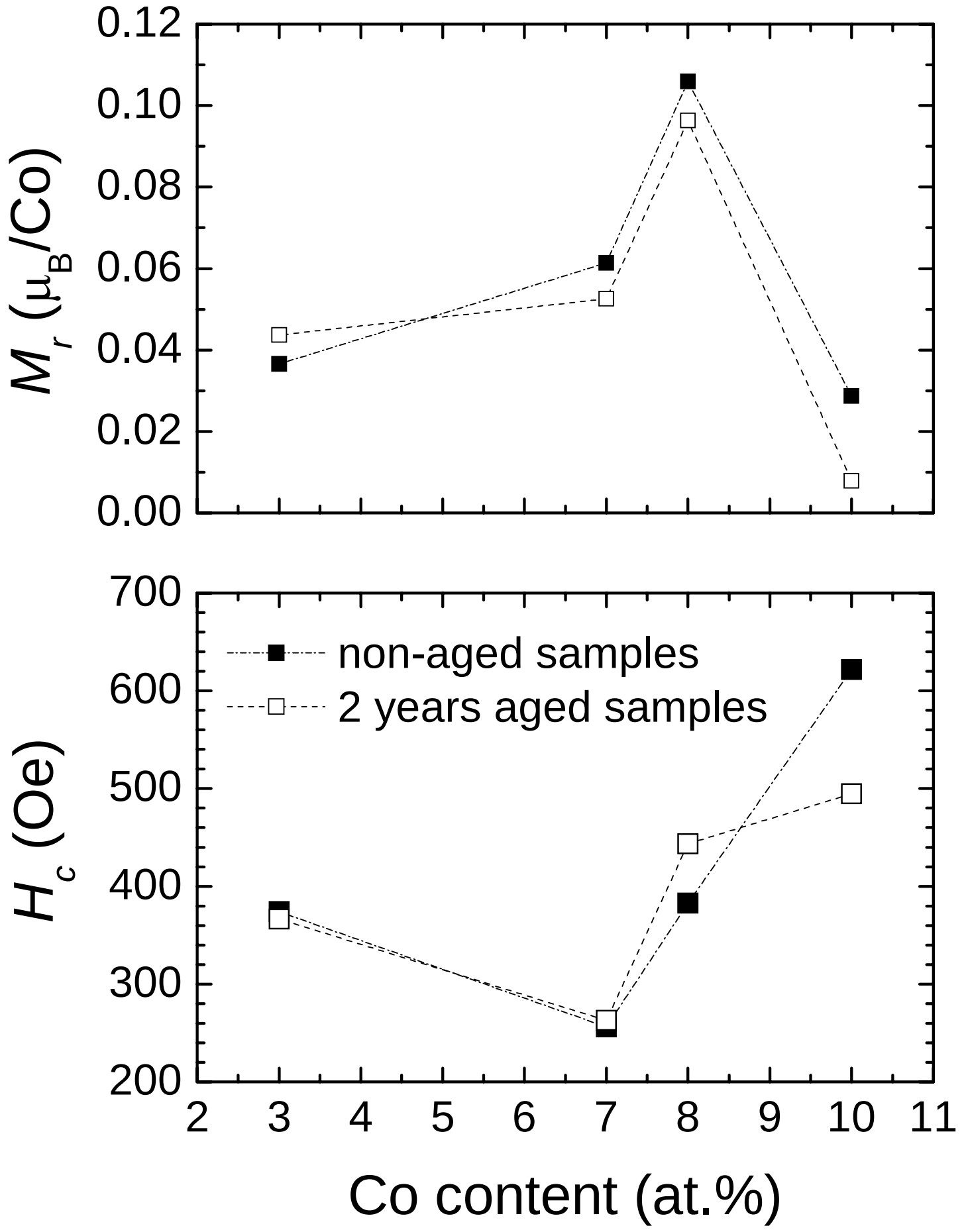

**Figure 4**

**Manuscript title:** Ferromagnetic order in aged Co-doped $TiO_2$ anatase nanopowders

**Authors:** A.J Silvestre *et al.*

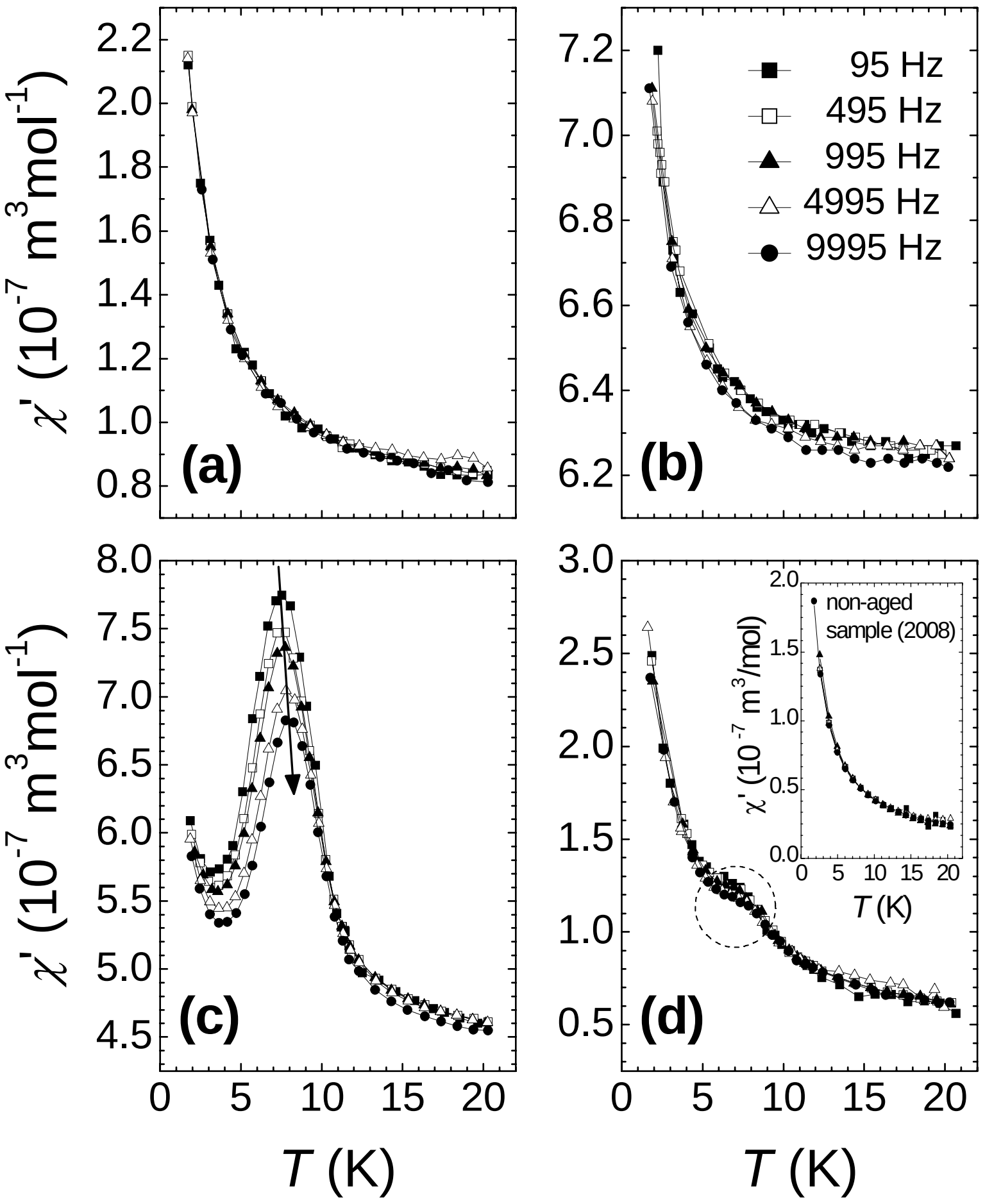

**Figure 5**

**Manuscript title:** Ferromagnetic order in aged Co-doped $TiO_2$ anatase nanopowders

**Authors:** A.J Silvestre *et al.*